# Diagnosis of COVID-19 based on Chest Radiography

Mei Gah Lim[1, a)] and Hoi Leong Lee[1, b)]

[1]*Faculty of Electronic Engineering Technology, Universiti Malaysia Perlis, Pauh Putra Campus, 02600 Arau, Perlis, Malaysia.*

[a)] Corresponding author: meigah115@gmail.com
[b)]hoileong@unimap.edu.my

**Abstract.** The Coronavirus disease 2019 (COVID-19) was first identified in Wuhan, China, in early December 2019 and now becoming a pandemic. When COVID-19 patients undergo radiography examination, radiologists can observe the present of radiographic abnormalities from their chest X-ray (CXR) images. In this study, a deep convolutional neural network (CNN) model was proposed to aid radiologists in diagnosing COVID-19 patients. First, this work conducted a comparative study on the performance of modified VGG-16, ResNet-50 and DenseNet-121 to classify CXR images into normal, COVID-19 and viral pneumonia. Then, the impact of image augmentation on the classification results was evaluated. The publicly available COVID-19 Radiography Database was used throughout this study. After comparison, ResNet-50 achieved the highest accuracy with 95.88%. Next, after training ResNet-50 with rotation, translation, horizontal flip, intensity shift and zoom augmented dataset, the accuracy dropped to 80.95%. Furthermore, an ablation study on the effect of image augmentation on the classification results found that the combinations of rotation and intensity shift augmentation methods obtained an accuracy higher than baseline, which is 96.14%. Finally, ResNet-50 with rotation and intensity shift augmentations performed the best and was proposed as the final classification model in this work. These findings demonstrated that the proposed classification model can provide a promising result for COVID-19 diagnosis.

## INTRODUCTION

The novel Coronavirus disease 2019, or COVID-19, was discovered in Wuhan, China, in early December 2019 and has now expanded worldwide, becoming a pandemic [1]. The COVID-19 pandemic had already resulted in about 6.4M mortalities and 560M cases of COVID-19 infection in early July 2022 [2]. This pandemic is caused by the severe acute respiratory syndrome coronavirus 2 (SARS-CoV-2) infecting people and can be spread from person to person. Fever, cough, difficulty breathing and invasive lesions on both lungs are the symptoms of COVID-19 infections [3], [4]. When someone sneezes or coughs, the virus spreads through the air in the form of saliva droplets. Early detection of this disease can help monitor the health status of infected patients by timely isolating them for two weeks or even longer to receive immediate treatment to mitigate the unstoppable of the disease spreading [5], [6].

The gold standard for diagnosing COVID-19 pneumonia is reverse transcriptase-polymerase chain reaction (RT-PCR) test, it discovers the existence of SARS-CoV-2 in respiratory samples collected via nasal swabs. Due to this pandemic outbreak, many developing countries face a significant challenge as there are no sufficient RT-PCR testing kits and resources in hospitals. So, to test the samples, they have to collect and send the test samples to more advanced medical centers, creating massive testing for COVID-19 [7]. The gold standard RT-PCR test is highly specific, and it needs time to get the test result due to its laborious and complicated manual process [6]. Also, it requires extraction skills during nasal swab sampling to avoid from getting false-negative result [8].

Instead of using the RT-PCR test, radiography examination is an alternative method that can be used for COVID-19 screening where chest imaging like chest X-ray (CXR) and computed tomography (CT) is conducted and interpreted by doctors or radiologists to observe and identify the visual indicators caused by SARS-Cov-2 infection. This is an applicable technique for detecting COVID-19 cases by capturing the radiographic abnormalities such as ground-glass opacity (GGO) and lung consolidation [6]. However, the features visible on radiography images are only relatively similar to those seen by the human eye [9]. Even the expert radiologist has a difficulty distinguishing between COVID-19 pneumonia and other pneumonia such as viral pneumonia [7]. This is one of the biggest bottlenecks the radiologist faced as diagnosing COVID-19 cases from chest imaging is challenging and time-consuming. Moreover, the problem faced when analysing the COVID-19 cases from the chest radiography images includes the inter-observer and intra-observer variability in the diagnostic result [10]. It is because of the difficulty to differentiate between lung pneumonia diseases since they have similar radiographical patterns.

So, Artificial Intelligence (AI)-based or computer-aided methods are important tools that can be utilised to automatic diagnose COVID-19 cases objectively by performing the image classification task [11]. Referring to several studies, they reviewed that for COVID-19 patients, the GGO mostly affects the lower lobes compared to the middle lobes, and least affects the upper lobes [12], [13]. Additionally, [12] stated that the GGO is more frequent in right lower lobe than left lower lobe. This may be the major way to differentiate COVID-19 pneumonia from other pneumonia since GGO and lung consolidation are the common features of COVID-19 that can be extracted and learnt when performing image classification using deep learning (DL) methods [14].

Therefore, the main contributions in this study are summarised below:
1. To develop an automatic method based on a deep convolutional neural network (CNN) to assist radiologists in diagnosing COVID-19 cases from CXR.
2. To compare the performance of CNN models, including modified VGG-16, ResNet-50 and DenseNet-121 to classify CXR images into normal, COVID-19 pneumonia and viral pneumonia cases.
3. To evaluate the effect of image augmentation on the performance of CNN models for CXR image classification.
4. To validate a deep CNN method on CXR images acquired from patients for COVID-19 detection.

The remaining paper is organised as follows: Section 2 presents the research and previous works related to the field of deep learning approach to detect COVID-19 based on chest images. Besides, Section 3 describes the methodology of our study. In Section 4, the results and discussion are highlighted. Lastly, Section 5 is the conclusion and recommendations of future work of our study.

## LITERATURE REVIEW

In order to fight with the COVID-19 disease, there are several researchers had contributed their studies about the automatic screening of COVID-19 on chest images. Khan et al. [4] developed a deep CNN model, named as CoroNet based on Xception architecture to identify COVID-19 infection automatically based on CXR images. They compared the accuracy of 2,3, and 4-class classification, and CoroNet achieved 99%, 95% and 89.6% respectively. COVID-Net is a DL model proposed by Wang et al. [6] based on residual architecture. COVID-Net was built with human-machine collaborative technique and trained with their proposed dataset, called COVIDx with 13,975 CXR images to perform 3-class classification and obtained an overall test accuracy with 93.3%. Besides, Chowdhury et al. [15] utilised a DL-based transfer learning method to automatically distinguish between the CXR images into three categories. They compared the model performance between binary and multi-class classification. Also, they compared the model performance with and without image augmentation. Lastly, the accuracy for binary classification is greater than the multi-class classification, and the model trained with augmented dataset performed the best.

To deal with the COVID-19 unbalanced data problem, Sakib et al. [7] introduced an adaptive data augmentation algorithm by combining a generative adversarial network (GAN) model with multiple generic augmentation techniques to increase the number of COVID-19 samples. Nefoussi et al. [16] balanced the sample size for each class since there will be a bias in major class if using this unbalanced dataset. Moreover, due to insufficient data, [4], [17], [18] pretrained their model with ImageNet to prevent overfitting issue.

Furthermore, there are several researchers applied image augmentation on the dataset to expand the dataset size and the variety of dataset to reduce the overfitting problem [16], [19]. There are two types of augmentation, which are offline augmentation and online (on-the-fly) augmentation. A review of previous work from Joseph et. al [20] that measured the performance of the two methods had revealed that the accuracy of the augmentation on-the-fly method is greater than the online one. Also, this method can save up the disk memory as it only return the random augmented data at each iteration while training without adding it to the original dataset to increase the dataset size. Aggarwal et al. [21] stated that data augmentation can increase the variety of training images and become more likely to tackle the real condition at which the CXR images are taken from different imaging protocols.

According to [19], [22], [23], skip connections used in ResNet can prevent diminishing gradients or backpropagation errors. Also, they stated that the model with skip connections is better than a standard or traditional CNN. Moreover, DenseNet is the extension of ResNet with skip connections that may had a better results than ResNet as it provide maximum gradient flow in the network [24]. The standard CNN is shown in Figure 1 and skip connections in ResNet and DenseNet are shown in Figure 2 and Figure 3 respectively.

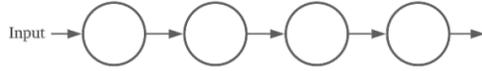

**FIGURE 1.** Standard CNN concept

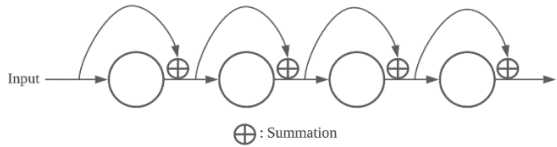

**FIGURE 2.** Skip connections in ResNet

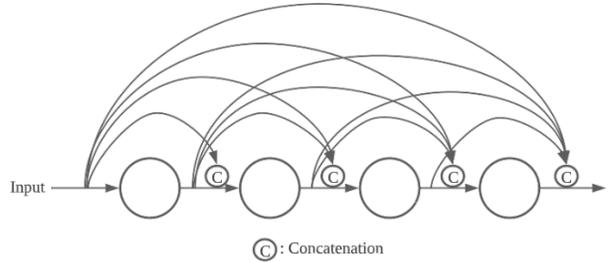

**FIGURE 3.** Skip connections in DenseNet

## METHODOLOGY

This section explained the pipeline of our study. Figure 4(a) is the flowchart for training and validation phase and Figure 4(b) is the flowchart for testing phase. It is started with the dataset preparation. The dataset is downloaded from an open access initiative. Then, the dataset is splitted into training, validation and testing set. The dataset is pre-processed and is ready to use to train the model for image classification. After that, the CXR images from testing set are fed into the trained model for evaluation.

Besides, we compared the model performance between three different CNN models (modified CNN, ResNet-50 and DenseNet-121) to classify the CXR images into 3-class. Also, we evaluated the impact of image augmentation on the performance of CNN model for CXR image classification. We implemented the augmentation step to the training and validation set after the pre-processing step.

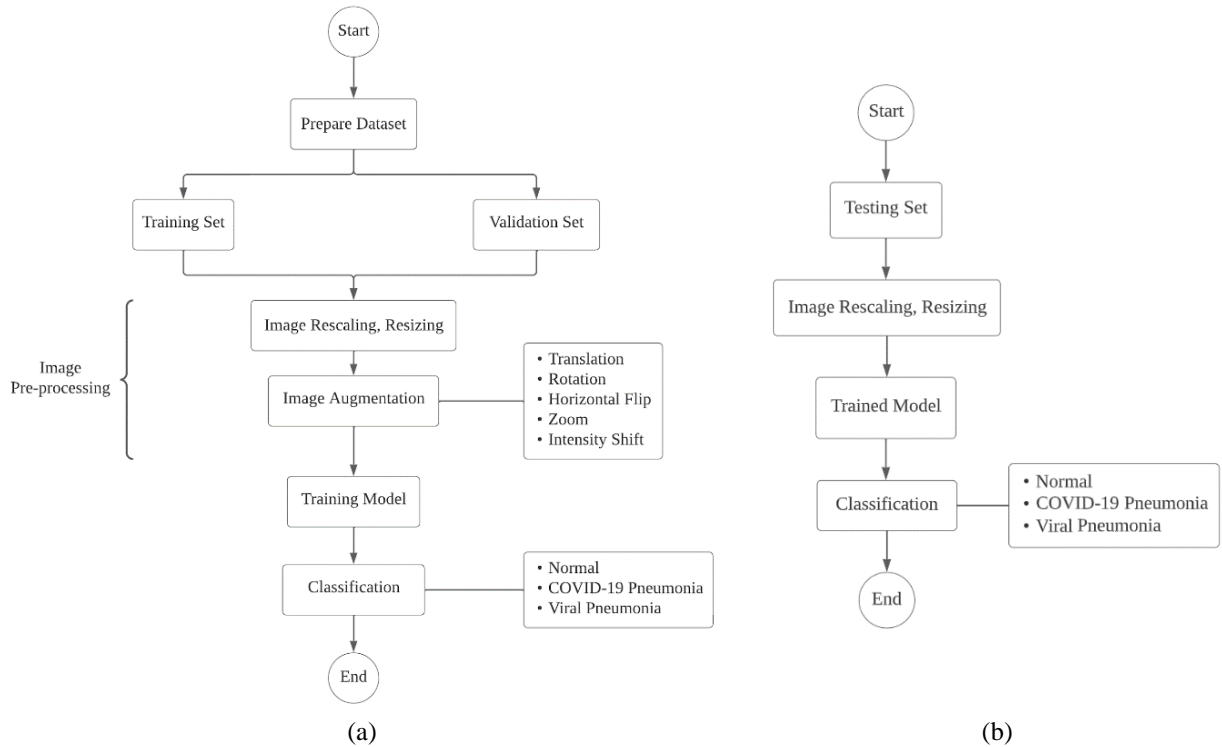

**FIGURE 4.** The flowchart for (a) Training and validation phase, and (b) Testing phase

## Model Architecture

The model architecture of modified VGG-16, ResNet-50 and DenseNet-121 are shown in Figure 5, Figure 6 and Figure 7 respectively. In this study, we modified the VGG-16 by reducing its number of filter by half, from 64 to 32 to reduce the network capacity and hence reduce overfitting. Besides, we used the original architecture of ResNet-50 and DenseNet-121.

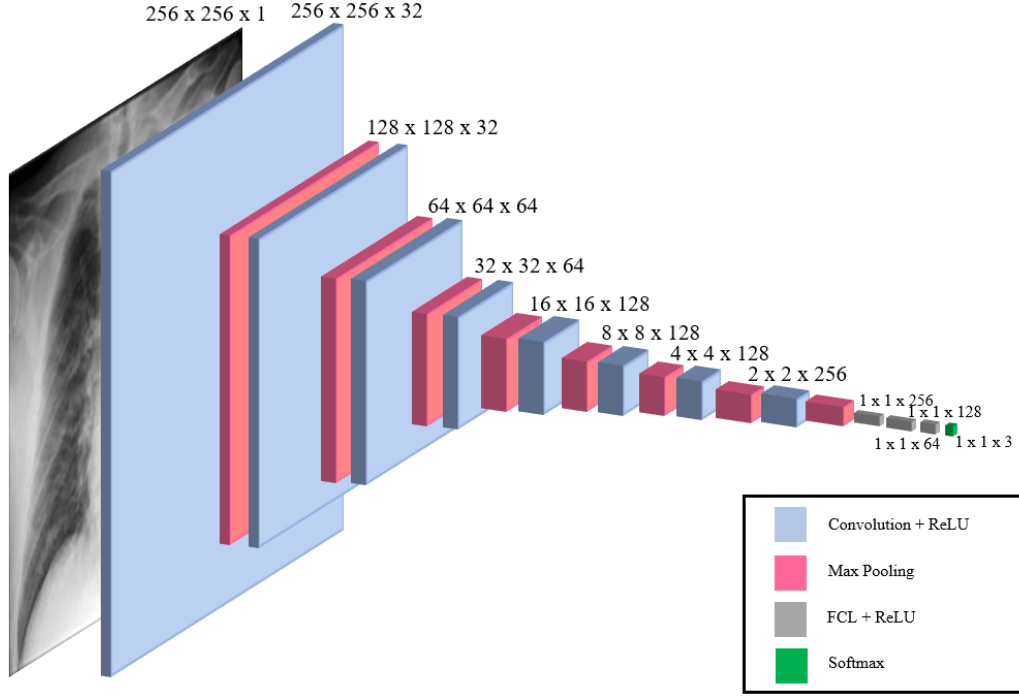

**FIGURE 5.** Modified VGG-16 architecture [25]

| Layer Name | Output Size | ResNet-50 |
|---|---|---|
| conv1 | 128×128 | 7×7, 64, stride 2 |
| conv2_x | 64×64 | 3×3 max pool, stride 2<br>$\begin{bmatrix} 1 \times 1, 64 \\ 3 \times 3, 64 \\ 1 \times 1, 256 \end{bmatrix} \times 2$ |
| conv3_x | 32×32 | $\begin{bmatrix} 1 \times 1, 128 \\ 3 \times 3, 128 \\ 1 \times 1, 512 \end{bmatrix} \times 3$ |
| conv4_x | 16×16 | $\begin{bmatrix} 1 \times 1, 256 \\ 3 \times 3, 256 \\ 1 \times 1, 1024 \end{bmatrix} \times 5$ |
| conv5_x | 8×8 | $\begin{bmatrix} 1 \times 1, 512 \\ 3 \times 3, 512 \\ 1 \times 1, 2048 \end{bmatrix} \times 2$ |
|  | 1×1 | average pool, 3D fully-connected, Softmax |
| FLOPs |  | $3.8 \times 10^9$ |

**FIGURE 6.** ResNet-50 architecture [23]

| Layers | Output Size | DenseNet-121 |
|---|---|---|
| Convolution | 128×128 | 7×7 conv, stride 2 |
| Pooling | 64×64 | 3×3 max pool, stride 2 |
| Dense Block (1) | 64×64 | $\begin{bmatrix} 1 \times 1 \text{ conv} \\ 3 \times 3 \text{ conv} \end{bmatrix} \times 6$ |
| Transition Layer (1) | 64×64 | 1×1 conv |
|  | 32×32 | 2×2 average pool, stride 2 |
| Dense Block (2) | 32×32 | $\begin{bmatrix} 1 \times 1 \text{ conv} \\ 3 \times 3 \text{ conv} \end{bmatrix} \times 12$ |
| Transition Layer (2) | 32×32 | 1×1 conv |
|  | 16×16 | 2×2 average pool, stride 2 |
| Dense Block (3) | 16×16 | $\begin{bmatrix} 1 \times 1 \text{ conv} \\ 3 \times 3 \text{ conv} \end{bmatrix} \times 24$ |
| Transition Layer (3) | 16×16 | 1×1 conv |
|  | 8×8 | 2×2 average pool, stride 2 |
| Dense Block (4) | 8×8 | $\begin{bmatrix} 1 \times 1 \text{ conv} \\ 3 \times 3 \text{ conv} \end{bmatrix} \times 16$ |
| Classification Layer | 1×1 | 7×7 global average pool<br>3D fully-connected, Softmax |

**FIGURE 7.** DenseNet-121 architecture [24]

## Dataset

For dataset preparation, we downloaded a publicly available dataset, COVID-19 Radiography Database [26] from Kaggle. This dataset contains 3-class (normal, COVID-19 pneumonia and viral pneumonia) with a total of 3886 CXR images and unbalanced number of images of each class. We splitted the dataset into train and test sets in the 8:2 ratio. Then, the train set was further divided in the 8:2 ratio for model training and validation. We applied random resampling method to balance the number of images for each class to avoid the classification result bias towards the major class. Table 1 displays the dataset summary.

**TABLE 1.** Dataset summary

| Type of Disease | Train | Validation | Test | No. of Images |
|---|---|---|---|---|
| Normal | 752 | 188 | 259 | 1199 |
| COVID-19 Pneumonia | 752 | 188 | 259 | 1199 |
| Viral Pneumonia | 752 | 188 | 259 | 1199 |
| Total | 2256 | 564 | 777 | 3597 |

## Image Pre-processing

Before loading the dataset for model training, validation and testing, the data images was undergoing a minimal pre-processing. During the pre-processing stage, the CXR images were rescaled or normalised to the range of [0,1] to make sure all the images are in the same manner and resized to 256×256 dimensions to make sure all the input images have a uniform size. Figure 8(a) and (b) depict the original CXR image and the image after pre-processed.

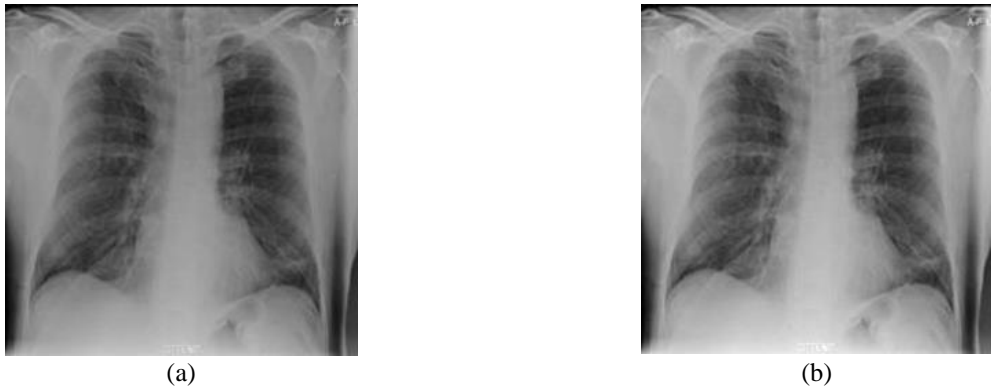

(a)      (b)

**FIGURE 8.** A sample of COVID-19 CXR image: (a) Original image, and (b) Pre-processed image

## Image Augmentation

Next, we evaluate the effect of image augmentation by comparing the testing accuracy of the CNN model when implementing the image augmentation to the training dataset. The types of augmentation applied to the dataset are shown in Table 2 and each augmentation type is visualised in Figure 9(a) to Figure 9(e). This augmentation step was proposed by Wang et al. [6] that is commonly utilised to the CXR images. Moreover, the examples of the CXR images from the dataset after implementing the combination of all of the stated augmentation techniques are shown in Figure 10. The Keras ImageDataGenerator was used to augment the CXR images on the fly. As to consider the clinical perspective and the augmentation that is applied to the dataset should not be aggressive, a small augmentation scale was used.

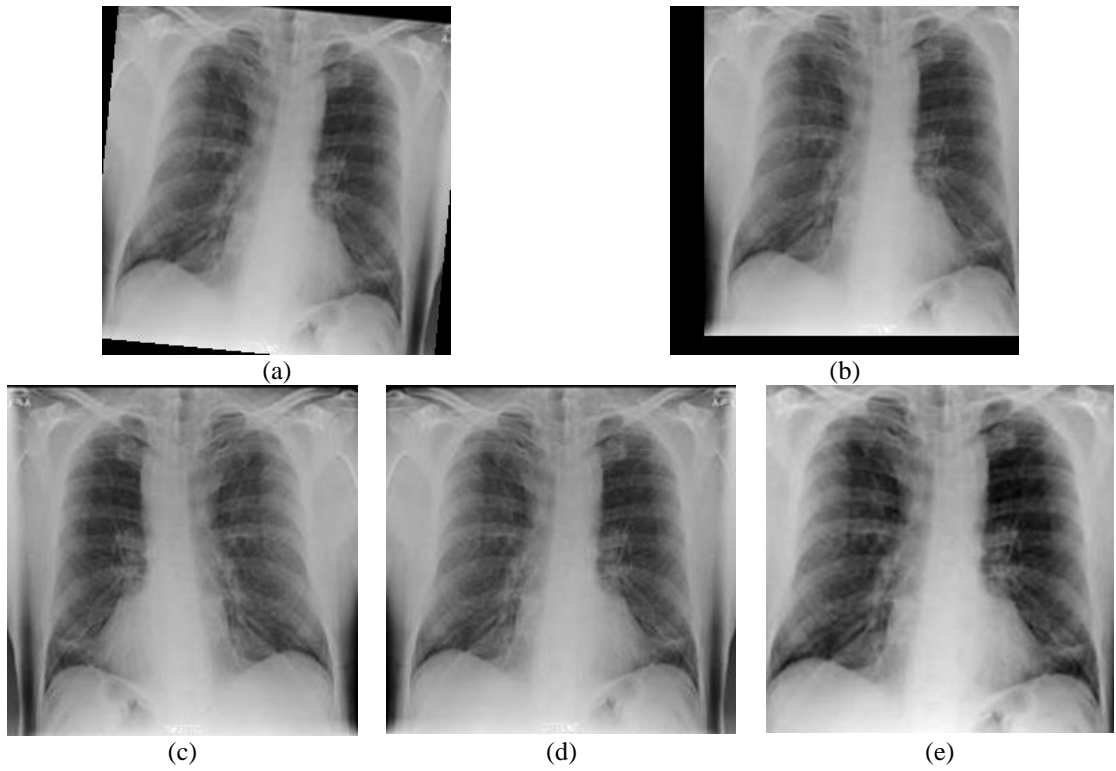

**FIGURE 9.** A sample of COVID-19 CXR image after augmented with (a) Rotation, (b) Translation, (c) Horizontal flip, (d) Intensity shift, and (e) zoom augmentation techniques.

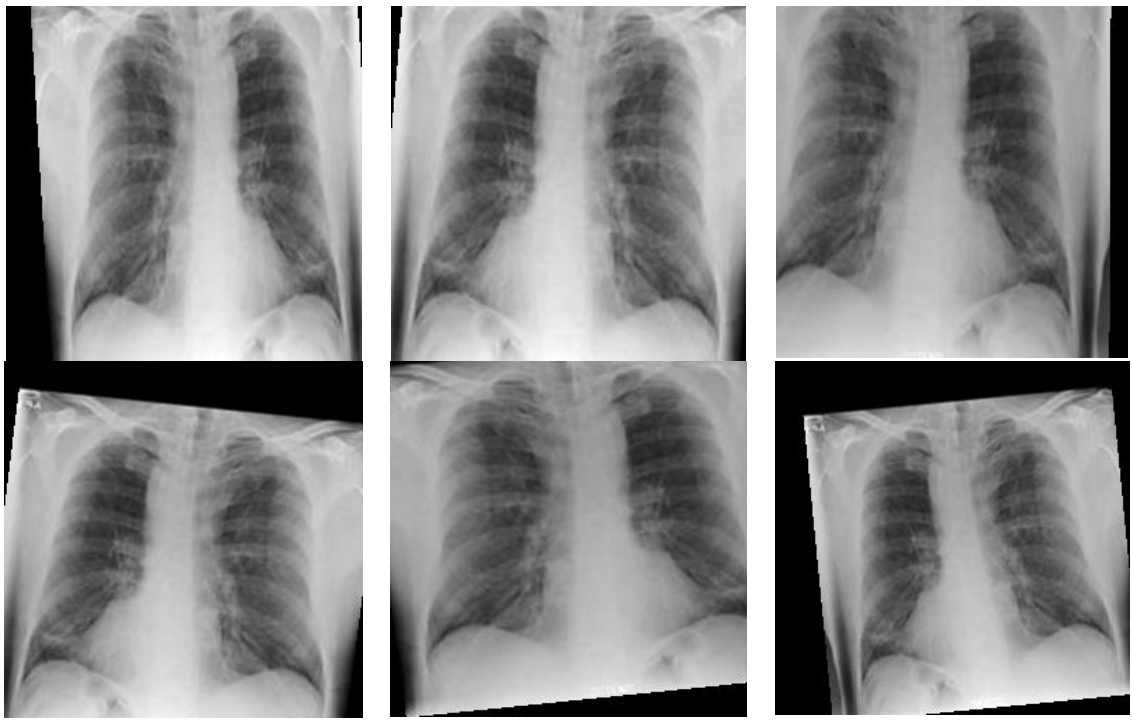

**FIGURE 10.** The results of random combined image augmentation techniques

## Training and Implementation Details

In this study, the proposed model was able to detect COVID-19 infections from CXR images automatically. The proposed model was constructed and trained to perform multi-class classification i.e. classify the CXR images into normal cases, COVID-19 pneumonia cases and viral pneumonia cases. The proposed model was implemented in Keras with TensorFlow as the backend. We trained the model end-to-end on the dataset by using the Adam optimizer [27] with hyperparameters as shown in Table 3. We used Google Colaboratory to conduct the experiment as it provided free access cloud-based GPU.

TABLE 2. Types of augmentation

| Augmentation Types | Description |
|---|---|
| Rotation | ± 10° |
| Translation | ± 10% in x and y direction |
| Horizontal Flip | True |
| Intensity Shift | ± 10% |
| Zoom | ± 15% |

TABLE 3. Training parameters

| Hyperparameters | Values |
|---|---|
| Learning Rate | 0.001 |
| Number of Epoch | 30 |
| Batch Size | 32 |
| Seed | 10 |

## RESULTS AND DISCUSSION

We carried out two experiments in this study. For the first experiment, we compared the model performance between different CNN models, which are modified VGG-16, ResNet-50 and DenseNet-121. For the second experiment, we evaluated the effect of image augmentation on the CNN model for image classification. Then, we proposed the model that performed the best throughout this study. We used the same hyperparameters as presented in Table 3 to fairly compare the classification result of the models.

### Experiment I: Comparison between Modified VGG-16, ResNet-50 and DenseNet-121

The CXR images from the dataset were pre-processed and then fed into the model for training, validation and testing. Table 4 shows the results summary of the comparison between modified VGG-16, ResNet-50 and DenseNet-121. From the result, ResNet-50 had achieved the highest test accuracy followed by modified VGG-16 and DenseNet-121.

The more parameters used, the more accurate it is. ResNet-50 has the largest number of parameters among these three CNN models. This may be the reason that ResNet-50 achieved the highest accuracy. Besides, from Table 4, we can see that the number of parameters of DenseNet-121 is more than that of modified VGG-16. However, when we compared between the accuracy of modified VGG-16 and DenseNet-121, the modified VGG-16 was more accurate than DenseNet-121. This probably due to the complexity of DenseNet-121 (121 layers) as it is too deep compared to the modified VGG-16 (16 layers), which also can be represented by the required training time. In addition, when the number of parameters increases, the training duration also increases as the computational complexity increases as seen in Table 4.

When the number of layers increase, the model becomes deeper. According to [24], a deeper model may have a better performance if the model is trained with a larger dataset. But, we used a relatively small training dataset in this study which cannot fit DenseNet-121 with 121 layers deep.

TABLE 4. Results summary for Experiment I

| CNN Models | Number of Parameters | Test Accuracy (%) | Training Time (s) |
|---|---|---|---|
| Modified VGG-16 | 770,531 | 94.47 | 1857.776 |
| ResNet-50 | 23,679,747 | 95.88 | 3081.350 |
| DenseNet-121 | 7,040,195 | 92.15 | 2009.301 |

### Experiment II: Impact of Image Augmentation on ResNet-50

In this part, we further evaluate the performance of ResNet-50 when trained with the augmented dataset. This experiment is aimed to evaluate the effect of image augmentation on the performance of the ResNet-50 model on the classification task. The CXR images from the dataset were pre-processed and then augmented before being fed into the ResNet-50 model for training, validation and testing. From the result shown in Table 5, ResNet-50 had achieved 80.95% test accuracy after training the model using an augmented CXR dataset. When compared to the results of the baseline model, the model trained with the dataset after applying image augmentation dropped by 14.93%. According to Elgendi et al. [28], image augmentation may have a bad effect on training phase as it creates noise to the image. So, we conducted an ablation study on ResNet-50 with image augmentation to evaluate the result of image augmentation.

*Ablation Study*

Table 7 shows the effect on the performance of the ResNet-50 model in terms of test accuracy when applying all possible combinations of image augmentation strategies to the dataset. The blue, red and black bolded numbers indicated the test accuracies of the baseline model (moderate), model that used all augmentations (lowest) and model that used only rotation and intensity shift augmentations (highest) respectively.

Besides, we can observe that when the model was trained with the dataset that applied the augmentation method individually, all the test accuracies were increased when compared to the one that applied all the augmentation methods together. Also, the model achieved the lowest test accuracy when training with the translation augmented dataset, and the model reached the highest test accuracy after training with the dataset with horizontal flip augmentation, followed by intensity shift augmentation. These two augmentation strategies yielded the results approximate to the baseline model, which is the model trained without image augmentation.

According to Table 7, it can be observed that only used a combination of augmentation methods, which is the combination of rotation and intensity shift augmentation methods, the ResNet-50 model can obtain the best performance. When we applied the other combinations to the dataset, the model test accuracy had improved when compared to the one that applied the combination of all augmentations together. However, these accuracies still did not exceed the test accuracy of the baseline model. Since most of the images in the "COVID-19 Radiography Database" are having intensity and rotation angle variation. So, this may be the reason that the combination of rotation and intensity shift augmentation methods scored the higher accuracy that the one without augmentation. This augmentation step helps the model to learn the variations in this dataset.

Table 5 shows the classification results when the ResNet-50 is trained with the dataset augmented with the combination of rotation and intensity shift augmentation methods. By comparison, this combination of augmentation methods improved the model test accuracy by 0.26%. Since this model had enhanced its performance after using the combination of rotation and intensity shift augmentation methods, we proposed this model as the classification model for COVID-19 detection.

However, this model only outperformed the baseline model by 0.26% accuracy. It can be said that the implementation of image augmentation may not influence the model performance for this dataset and this result may only be applicable to this study. Besides, Nigam et al. [29] mentioned that the application of image augmentation in medical imaging field is not considered as a good practice. The work done by Chowdhury et al. [15] revealed that the models with an increased classification accuracy after implementing image augmentation but the difference is small, not more than 1.5%. Also, the classification accuracy of some models dropped insignificantly with image augmentation, not exceeding 2%.

Furthermore, the result in this study is similar to the result obtained by Elgendi et al. [28] who used three different publicly available datasets with four different combinations of augmentation methods to evaluate the effect of data augmentation on the model performance for COVID-19 detection. None of the augmentation methods improves the model performance. Additionally, Elgendi et al. stated that many researchers always assume that the model performance will improved after applying data augmentation, however, it depends on the case or dataset and the used of augmented techniques.

Table 6 presents the indirect comparison of the final proposed classification model in this study with the other DL models with different state-of-the-art that had been proposed by the other authors. Hence, this proposed model is able to provide a better or comparable classification result.

**TABLE 5.** Result summary of Experiment II

| Augmentation Types | Test Accuracy (%) |
| --- | --- |
| None (Baseline) | 95.88 |
| All | 80.95 |
| Rotation + Intensity Shift | 96.14 |

**TABLE 6.** Indirect comparison of the proposed model with other DL models

| Study | Overall Accuracy (%) |
| --- | --- |
| CoroNet [4] | 95.00 |
| COVID-Net [6] | 93.30 |
| Proposed Model | 96.14 |

TABLE 7. ResNet-50 testing results after applied all possible combinations of augmentation techniques

| Augmentation Types | Rotation | Translation | Horizontal Flip | Intensity Shift | Zoom | Test Accuracy (%) |
|---|---|---|---|---|---|---|
| None | | | | | | **95.88** |
| All | ■ | ■ | ■ | ■ | ■ | **80.95** |
| Individual | ■ | | | | | 94.08 |
| | | ■ | | | | 92.54 |
| | | | ■ | | | 95.75 |
| | | | | ■ | | 95.62 |
| | | | | | ■ | 94.77 |
| 2 Combined | ■ | ■ | | | | 93.05 |
| | ■ | | ■ | | | 93.69 |
| | ■ | | | ■ | | **96.14** |
| | ■ | | | | ■ | 89.58 |
| | | ■ | ■ | | | 89.19 |
| | | ■ | | ■ | | 95.62 |
| | | ■ | | | ■ | 89.19 |
| | | | ■ | ■ | | 95.11 |
| | | | ■ | | ■ | 94.34 |
| | | | | ■ | ■ | 90.22 |
| 3 Combined | ■ | ■ | ■ | | | 92.92 |
| | ■ | ■ | | ■ | | 92.02 |
| | ■ | ■ | | | ■ | 88.67 |
| | | ■ | ■ | ■ | | 92.79 |
| | | ■ | ■ | | ■ | 91.89 |
| | | | ■ | ■ | ■ | 92.66 |
| | ■ | | ■ | ■ | | 95.97 |
| | | ■ | | ■ | ■ | 94.08 |
| 4 Combined | ■ | ■ | ■ | ■ | | 92.79 |
| | ■ | ■ | ■ | | ■ | 92.41 |
| | ■ | ■ | | ■ | ■ | 93.56 |
| | ■ | | ■ | ■ | ■ | 91.89 |
| | | ■ | ■ | ■ | ■ | 93.95 |

## Performance Evaluation

After testing the model, we evaluate the model performance in terms of accuracy, precision, recall and F1-score. The mathematical equations are listed as below. Table 8 presents the classification report of our proposed classification model, which is the ResNet-50 with rotation and intensity shift augmentation methods.

- Accuracy: determine the ability of the model to classify the three cases correctly.

$$\text{Accuracy} = \frac{TP + TN}{TP + FP + TN + FN} \quad (1)$$

- Precision: determine the portions of the true predictions among all the predictions.

$$\text{Precision} = \frac{TP}{TP + FP} \quad (2)$$

- Recall: determine the true positive cases which it already been diagnosed as positive.

$$\text{Recall} = \frac{TP}{TP + FN} \quad (3)$$

- F1-score: weighted average of precision and recall.

$$\text{F1-score} = 2 \times \frac{\text{Precision} \times \text{Recall}}{\text{Precision} + \text{Recall}} \quad (4)$$

where TP = True positive, TN = True negative, FP = False negative, and FN = False negative.

**TABLE 8.** Classification report of the proposed classification model

| Classes | Accuracy | Precision | Recall | F1-score |
|---|---|---|---|---|
| COVID-19 |  | 0.98 | 0.98 | 0.98 |
| Normal | 0.96 | 0.95 | 0.97 | 0.96 |
| Viral Pneumonia |  | 0.96 | 0.93 | 0.95 |

According to the confusion matrix shown in Figure 11, for a total number of 259 images for each category, 254, 251 and 242 images were predicted correctly as COVID-19, normal and viral pneumonia classes respectively. Also, there were 1 and 4 COVID-19 images misidentified as normal and viral pneumonia cases respectively. Next, for normal cases, only 1 image was predicted wrongly as COVID-19 and 7 images were misclassified as viral pneumonia cases. Then, 5 out of 259 viral pneumonia CXR images were misclassified as COVID-19 images.

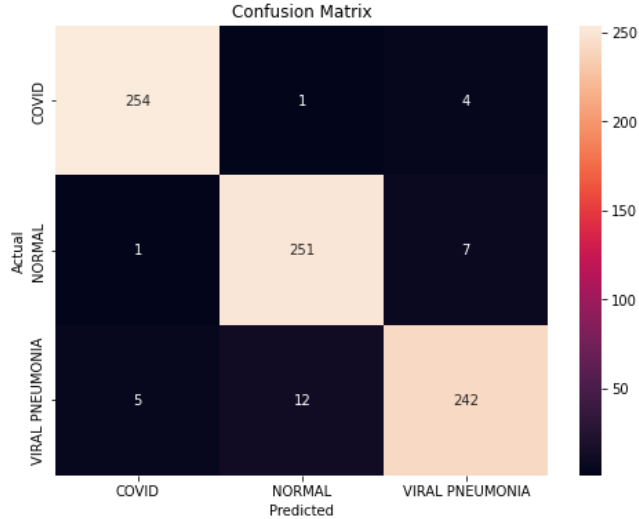

**FIGURE 11.** Confusion matrix of the proposed classification model

## CONCLUSION

In this work, we proposed an automatic approach for COVID-19 diagnostic based on CXR images. The method pipeline included a basic image pre-processing step, rotation and intensity shift image augmentation strategies and deep CNN. This approach has the potential to contribute to assisting radiologists in the diagnosis of COVID-19 as it can automatically classify COVID-19 infections from CXR images, which successfully achieved the main objective of this research. Our proposed model is able to perform multi-class classification by distinguishing between normal, COVID-19 pneumonia and viral pneumonia cases, and provide a promising result with 96.14% overall accuracy.

### Recommendation of Future Work

In order to improve the future work on diagnosing COVID-19 based on CXR images, there are several recommendations can be proposed. Firstly, we may include a study of CNN multi-class classification performance between transfer learning model and without transfer learning model to evaluate the importance of transfer learning techniques in increasing the classification accuracy of a small dataset.

In addition, we expand the size of the dataset in the future to increase the model performance. We enlarge the size of the dataset by combining various publicly available datasets to increase the variety of training images to avoid memorisation. This action will increase the model generalisation.

Furthermore, we suggest to apply various image enhancement pre-processing techniques to the dataset. For instance, the image pre-processing method such as histogram equalisation and CLAHE used to enhance the image contrast and improve the image quality are suggested for implementation in future work. From the previous work,

Al-Waisy et al. [19] stated that the contrast limited adaptive histogram equalization (CLAHE) method can improve the small details, textures and low image contrast of the CXR images. This method may improves the image local contrast and sharpen the image edges.

Last but not least, we are recommended to modify the ResNet-50 model architecture by adding dropout layers and convolutional layers to improve the accuracy and model performance. As mentioned in previous work, adding dropout layer can prevent overfitting problem [4]. According to Simonyan et al. [25], adding more convolutional layers can increase the depth and hence brings a better performance. For a deeper model, make sure that it is trained with larger dataset. For a large dataset, the learning rate can decrease so that the neural network can have sufficient time to learn the important features. By applying a learning rate scheduler, the learning rate will reduce by a factor if the validation loss stopped decreasing.

## ACKNOWLEDGMENTS

I want to thank my supervisor, Dr. Lee Hoi Leong, who has supported my research. He provided and shared valuable knowledge and guidelines on research skills.